\documentclass[%
 reprint,%
 amssymb, amsmath,%
 aip,cha,%
]{revtex4-1}

\usepackage{bm}%
\usepackage[colorlinks=true,linkcolor=blue]{hyperref}%
\usepackage{amsmath}
\usepackage{graphicx}
\usepackage{cancel}
\usepackage{subfigure}
\usepackage{mathrsfs}
\expandafter\ifx\csname package@font\endcsname\relax\else
 \expandafter\expandafter
 \expandafter\usepackage
 \expandafter\expandafter
 \expandafter{\csname package@font\endcsname}%
\fi
\begin{document}

\title{Machine-learning-assisted electron-spin readout of nitrogen-vacancy center in diamond}%

\author{Peng Qian}%
\author{Xue Lin}%
\author{Feifei Zhou}%
\author{Runchuan Ye}%
\author{Yunlan Ji}%
\author{Bing Chen}%
 \altaffiliation{\textbf{Authors to whom correspondence should be addressed:} nyxu@hfut.edu.cn, bingchenphysics@hfut.edu.cn}
\author{Guangjun Xie}%
\author{Nanyang Xu}%
 \altaffiliation{\textbf{Authors to whom correspondence should be addressed:} nyxu@hfut.edu.cn, bingchenphysics@hfut.edu.cn}
\affiliation{School of Electronic Science and Applied Physics,Hefei University of Technology, Hefei, Anhui 230009, China}%

\date{\today}%

\begin{abstract}
Machine learning is a powerful tool in finding hidden data patterns for quantum information processing. Here, we introduce this method into the optical readout of electron-spin states in diamond via single-photon collection and demonstrate improved readout precision at room temperature. The traditional method of summing photon counts in a time gate loses all the timing information crudely. We find that changing the gate width can only optimize the contrast or the state variance, not both. In comparison, machine learning adaptively learns from time-resolved fluorescence data, and offers the optimal data processing model that elaborately weights each time bin to maximize the extracted information. It is shown that our method can repair the processing result from imperfect data, reducing 7\% in spin readout error while optimizing the contrast. Note that these improvements only involve recording photon time traces and consume no additional experimental time, they are thus robust and free. Our machine learning method implies a wide range of applications in precision measurement and optical detection of states.
\end{abstract}
\date{\today}%
\maketitle

Machine learning (ML)  has been widely applied to various communities of quantum information science due to its outstanding data-mining capability. Many recent advancements benefit from this technique, including  identifying entanglement\cite{gao2018experimental,lu2018separability-entanglement,Ma:2018aa},
 enhancing quantum sensing\cite{Santagati_2019, Dinani_2019}, 
 reducing crosstalk in multi-qubit readout\cite{seif2018machine}, 
 simulating many-body systems\cite{Carleo_2017},
 and constructing decoders for quantum error correcting codes\cite{krastanov2017deep,torlai2017neural}. ML algorithms teach the computer to learn from noisy data and extract information to automatically build up a model. Therefore, this technique is especially suitable for fault-tolerant information processing, where highly accurate readout of quantum states from the noise environment is demanded\cite{bermudez2017assessing}.
  
For most qubit state discrimination tasks relying on photoluminescence photons, a detection time gate is defined to collect photons. Since traditionally only the total of counts is used to label a state, the inner features of the photon time trace are all deserted\cite{steiner2010universal}. This is suffered by the threshold method  employed in single-shot readout systems for classifying two state distributions\cite{neumann2010single}. During the long-time laser illumination for state detection, for example, in a trapped ion or a superconducting qubit system, unpredicted state flip can happen, which could have been revealed in the time trace\cite{ding2019fast, magesan2015machine}. Other qubit systems where very few photons are detected in a single measurement, like room-temperature nitrogen-vacancy (NV) center in diamond, require statistical readout other than single shot. Changing the gate duration and the delay between the time gate and the excitation laser pulse can  partly filter out the fluorescence background and thus improve the readout performance, which is yet very limited\cite{chen2019high}. However, the ML method is capable of fully exploiting the timing information of the photon trace, exploring hidden data patterns and providing the optimal solution to data processing\cite{liu2020repetitive}. Therefore, it is very suitable to apply ML to these experiments to maximize the readout performance.

\begin{figure}[htb]
\centering
\includegraphics[width=0.48\textwidth]{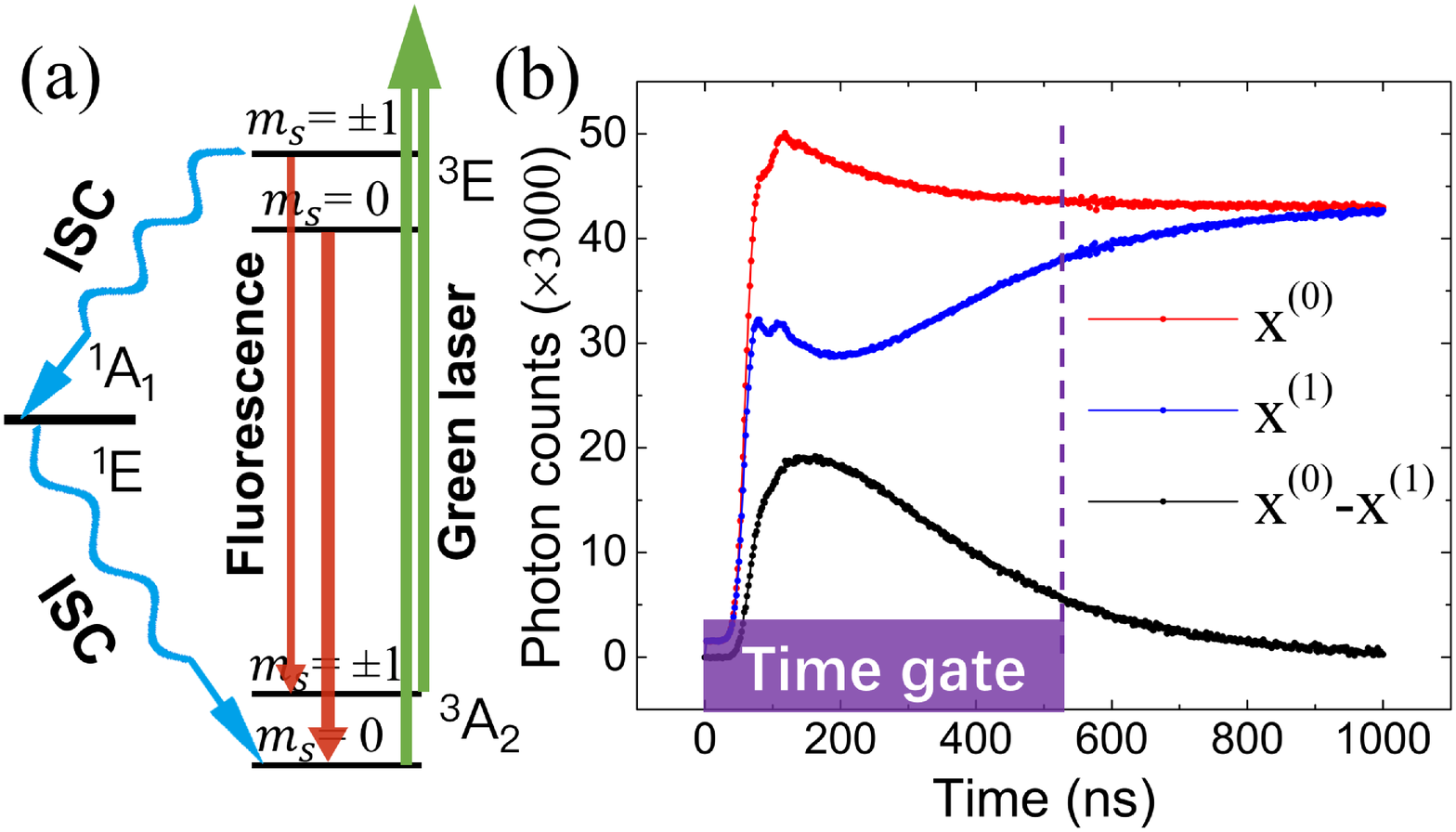}
\caption{(a) A simple scheme of NV center level structure. The ISC process transfers the spin into $m_s=0$ finally. (b) Photon time traces of spin states initialized on $m_s=0$ (red) and $m_s=1$(blue), measured with $10^9$ repetitions. The black curve is the differential signal of them. }
\label{fig:setup}
\end{figure}
     
 In this work, we focus on applying ML to the electron-spin readout of  room-temperature NV center in diamond\cite{doherty2013nitrogen, geng2016experimental, rong2015experimental}, aimed at improving the spin readout at low cost. A simple model of the NV center energy-levels is shown in Fig.\ref{fig:setup}(a). The transition between the ground triplet state $^{3}A_{2}$ and the excited triplet state $^{3}E$ enables optical addressability of NV centers with our homebuilt confocal microscopy system\cite{chen2020detecting}. The ground state is excited with a 532 nm laser, while the excited state releases fluorescence when it radiatively relaxes down to the ground state with the spin conserved\cite{fuchs2010excited}. However, the $m_{s}=\pm1$ state has a much higher probability than $m_{s}=0$ to non-radiatively cross over from the triplet into the singlet manifold, which is called intersystem-crossing (ISC)\cite{goldman2015phonon, Suter_2017}. After being shelved for about 250 ns on the metastable state $^{1}E$, the spin preferentially returns to $m_{s}=0$ of the ground state\cite{robledo2011spin}. In this way, the system is finally initialized in the $m_{s}=0$ state after several optical cycles\cite{chen2019quantum}. The key here is that, the $m_{s}=0$ state fluoresces more than $m_{s}=\pm1$ state in the first few hundred  nanoseconds, which leads to about 30\% more total counts in contrast. This mechanism enables optical readout of NV center spin states at room temperature, which is the foundation of NV-center-based quantum information processing. 
          
Ideally, researchers want pure negative charge state, perfect initialization, exact spin flipping\cite{xu2019dynamically}and high emission\cite{marseglia2011nanofabricated, jiang2009repetitive}. These could be easily affected by drifting factors like laser power, sample position, environment temperature, etc. This kind of system noise is beyond the scope of this article. Here, we focus on another main kind of  noise regarding merely the registered photons, namely shot noise, which comes from the particle nature of photons. The photodynamics inside the NV center are also complex enough to present randomness. We expect to suppress shot noise and maximize the usage of fluorescence signal with ML.
          
Typical fluorescence time traces are shown in Fig.\ref{fig:setup}(b). They are detected after the readout laser is turned on and accumulated with $10^9$ measurements, recorded with a self-made 2ns-resolution time tagger based on a commercial field-programmable-gate-array (FPGA) module. We set a time gate on the traces as usual that cuts off before the traces reach steady states, as illustrated by the dashed line.  The gate width $\Delta t$ is usually optimized for different NV centers\cite{steiner2010universal}. Suppose there are $N$ time bins in the gate, the traditional method just adds up the photon counts dwelling in each bin, i.e., $\sum x_{i}, i=1,2, ..., N$. For the two traces of polarized states $m_{s}=0$ and $m_{s}=1$, their total counts represent the lower and upper  fluorescence boundary of the NV system, implying 100\% and zero population on $m_{s}=0$ state, respectively. Any superposition of the two states generates photon counts between the two boundaries under the same measurement repetitions. If we normalize the photon counts to the population probability on $m_{s}=0$ as $p$ ($p\in [0,1]$), then the mapping from a time trace to this $p$ is supposed to be a linear combination of  all the time bins:
\begin{equation}
\label{eq:p}
p^{(\psi)}=\sum_{i=1}^{N}a_{i} x_{i}^{(\psi)}+b,
\end{equation}               
where $\psi$ represents any  superposition state  of $|m_{s}=0\rangle$ and $|m_{s}=1\rangle$, $\{a_{i}\}$ are linear coefficients and $b$ is a constant. Considering that there are only $\sim0.02$ photons on average collected within one measurement in our experiment,  it's reasonable to assume Poisson distribution within each bin. Then the variance of $p$ can be calculated as $\sigma_{p}^2=\sum_{i=1}^{N}a_{i}^2 x_{i}$, where $\sigma_p$ is the standard deviation.

\begin{figure}[htb]
\centering
\includegraphics[width=0.48\textwidth]{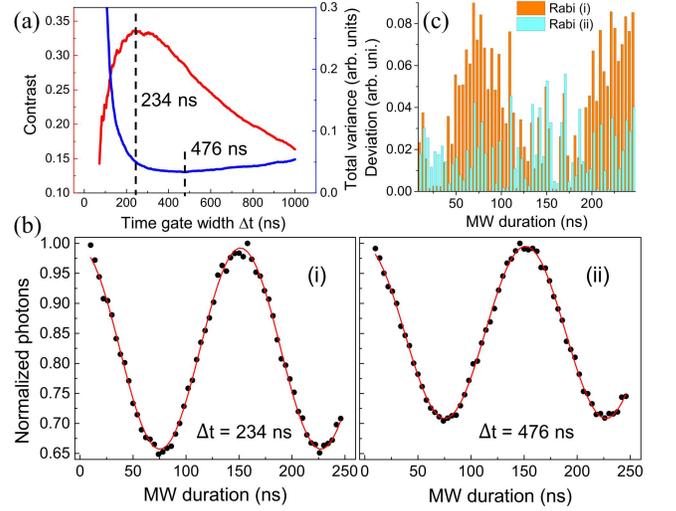}
\caption{ (a) Calculated contrast $\mathcal{C}$ and total variance $\mathcal{V}$ using the time traces of states $m_s=0$ and $m_s=1$ ($10^6$ measurements) with an increasing time gate width. The contrast $\mathcal{C}$ takes the maximum at 234 ns while the total variance $\mathcal{V}$ takes the minimum at 476 ns. (b) Resulted Rabi oscillations after processing a set of Rabi data of $10^6$ measurements with the two optimal gate widths in (a) for different metrics. (c) Columns of differences between the data points and the best fitted lines in (b) for clear vision. A general reduction in Rabi (ii) is seen compared with Rabi (i).}
\label{fig:timegate}
\end{figure}

\begin{figure}[htb]
\centering
\includegraphics[width=0.48\textwidth]{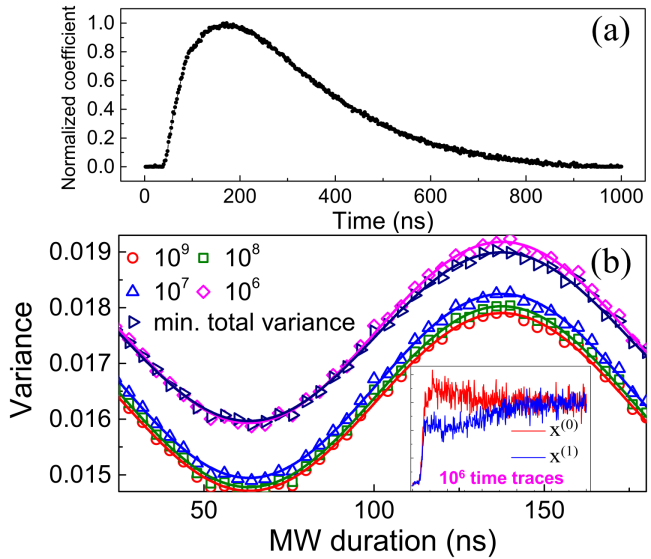}
\caption{ (a) A regression coefficient distribution over time bins trained with the two boundary traces in Fig.\ref{fig:setup}(b) . (b) Calculated variances on a test set of Rabi data of $10^5$ measurements with different coefficient models trained with boundary traces of $10^9, 10^8, 10^7, 10^6$ measurements, respectively. Their average variances are respectively about 6.8\%, 6.3\%, 5.2\% and -0.3\% lower than that of the mini-$\mathcal{V}$ metric, and are all over 40\% lower than that of the max-$\mathcal{C}$ metric (average value 0.0288, not shown in the figure due to being too far above). The solid curves are for eye's guide. Inset: time traces of $10^6$ measurements with large noise. }
\label{fig:01}
\end{figure}

Let's see how the traditional method works. Since the photons of time bins are directly summed, each $a_{i}$ has the same value denoted as $a$.  Given the two boundary conditions $p^{(0)}=1$ and $p^{(1)}=0$, it can be calculated that $a=1/(L_{0}-L_{1}), b=-L_{1}/(L_{0}-L_{1})$, where $L_{0}=\sum x_{i}^{(0)}, L_{1}=\sum x_{i}^{(1)}$. Then Eq.\ref{eq:p} becomes
\begin{equation}
\label{eq:ptimegate}
p^{(\psi)}=\frac{1}{L_{0}-L_{1}}\sum_{i=1}^{N} x_{i}^{(\psi)}-\frac{L_{1}}{L_{0}-L_{1}}.
\end{equation} 
This function describes the traditional way of using data, and the variance becomes $\sigma_{p}^2=1/(L_{0}-L_{1})^2\sum_{i=1}^{N} x_{i}$. In fact, in NV-center experiments, the most commonly used metric for deciding the time gate width is dimensionless signal contrast, which is defined as $\mathcal{C}=(L_0-L_1)/L_0$\cite{hopper2018spin}. To demonstrate how the choice of $\Delta t$ influences the readout contrast and variance, we respectively calculated these two metrics under various gate width conditions, using the time traces of polarized spin states $m_s=0$ and $m_s=1$ measured with $10^6$ repetitions (shown in the inset of  Fig.\ref{fig:01}(b)).  To reflect the whole effect of readout errors of all the possible states, we integrate $\sigma_{p}^2$ over the range $[0,1]$ as $\mathcal{V}=\int_0^1\sigma_{p}^2 dp=(L_0+L_1)/2(L_0-L_1)^2$, and call it total variance. As shown in Fig.\ref{fig:timegate}(a), the contrast $\mathcal{C}$ takes the maximum at $\Delta t$=234 ns, while the total variance $\mathcal{V}$ takes the minimum at $\Delta t$=476 ns. This indicates that different choices of gate width will lead to the optimization of different readout metrics. To visually verify this conclusion, we further processed the whole set of Rabi data of $10^6$ measurements including the two polarized states above, with the two gate widths $\Delta t$=234 and 476 ns, respectively, as shown in Fig.\ref{fig:timegate}(b). The contrast of  Rabi (i) reaches $\sim\text{33}\%$, while that of Rabi (ii) drops to $\sim\text{27}\%$, but the data points in Rabi (ii) are a little better aligned on a sinusoidal curve. For clear vision, Fig.\ref{fig:timegate}(c) displays the differences between the points and the best fitted lines in Rabi (i) and Rabi (ii). The general reduction in data deviation indicates a more accurate spin state readout. Therefore, our defined population variance is a good metric to evaluate the state readout precision. Changing the time gate alone in the traditional method cannot optimize the two metrics simultaneously.

However, it is not necessary to equally weight each time bin. The purpose of data processing is extracting as much information as possible  from data to facilitate discrimination of qubit states. It can be observed that some bins contribute much larger in differential fluorescence than others along the time axis. Allowing the coefficients \{$a_{i}$\} to differ from each other flexibly would strengthen the roles of some bins that have more positive effect on readout precision than negative. Ref\cite{gupta2016efficient} deals with a  similar issue, but its using maximum likelihood estimation consumes much time to acquire calibration data and needs assumptions to build up the analytic approach. In contrast, our ML method directly gives non-binary weights for time bins via algorithms and grasps relatively more information with fewer data. Besides, we discarded the cut-off time gate and make use of  the whole time trace, because the ML algorithms should be clever enough to treat each time bin appropriately.

The form of Eq.\ref{eq:p} visually implies the use of linear regression technique to build our prediction model. Technically, the parameters $\{a_{i}\}$ are referred to as regression coefficients. The linear regression works as follows: first, it predicts a $p$ given a set of input data $\{x_{i}\}$ according to the regression coefficient model; then, the predicted $p$ is compared with the target value (close to zero in a trough and to one in a peak of the Rabi oscillations, for instance); finally, the regression coefficients are updated towards reducing the distance between $p$ and the target. Therefore, the loss function can be expressed as $J(a_{i},b)=\sum_{j=1}^{m}(h(\boldsymbol{x})^{(j)}-q^{(j)})^2$, where $m$ is the number of training examples, the superscript $j$ denotes the $j$th training example, $\boldsymbol{x}$ is the vector representation of time-binned photons in a trace,  $h(\boldsymbol x)=p$ as in Eq.\ref{eq:p} and the target value is denoted as $q$. All the coefficients should be positive. Here the  gradient descent algorithm is used to repeatedly decrease $J(a_{i},b)$ and change ${a_i}$ and $b$ until some converge condition is reached, when the optimal coefficients are found. However, this loss function may lead to multiple results since the coefficients are not restricted at all. Hence, we physically propose to add one term $\sum\sigma_{p}^2$ to the function, namely the total variance $\mathcal{V}$ of all the training examples. The final loss function becomes
\begin{equation}
\label{eq:loss}
J(a_{i},b)=\sum_{j=1}^{m}(h(\boldsymbol{x})^{(j)}-q^{(j)})^2+\sum_{j=1}^{m}\sigma_{p}^2.
\end{equation} 
The two terms in Eq.\ref{eq:loss} are essentially contradictory, because reducing the second term will sacrifice the prediction accuracy of the first term. Still, minimizing the two terms simultaneously can balance the variance and the deviation from the target, which guarantees that the resulted $\{a_i\}$ won't go to extremes. This is the key of our ML method. In general, the linear regression can evaluate how strongly each time bin influences the spin readout, and is a popular tool of ML community.

Now we use the two boundary traces of $10^9$ measurements in Fig.\ref{fig:setup}(b) as a training set of two examples to test our method. They have target values $q^{(1)}=1, q^{(2)}=0$, respectively. In the algorithms, we give the first term of Eq.\ref{eq:loss} priority in optimization over the second term to guarantee that the predicted value of $h(\boldsymbol x)$ is as close as possible to the target $q$. This is achieved by multiplying the first term with a much larger weight factor than that for the second term. Because the first term decides how well our model characterizes the spin state population, the optimization of the spin readout error should not damage the contrast. The training set data are normalized before being fed to the loss function, which can speed up the converging of the model and raise the prediction accuracy. The resulted regression coefficient distribution is plotted in Fig.\ref{fig:01}(a). It has the same trend as the differential signal in Fig.\ref{fig:setup}(b), exactly according with what we expected: those time bins contributing more to the signal have larger weights. Then we applied this coefficient model to a set of Rabi data (test set) comprising of 60 test examples  each with $10^5$ measurements, and calculated their variances (red circles in Fig.\ref{fig:01}(b)). To compare, we also calculated the variances in the two traditional time-gated metrics. The state readout variance level is reduced about 7\% compared with that of the minimal-total-variance (mini-$\mathcal{V}$) metric, and 44\% compared with the maximal-contrast (max-$\mathcal{C}$) metric.  One might think that  $10^9$ measurements for obtaining a training set cost too much time. So we further acquired three boundary training sets of $10^8, 10^7, 10^6$ measurements, and applied them to the same test set. Although the time traces of $10^6$ measurements have large noise (in the inset of  Fig.\ref{fig:01}(b)), they still lead to a readout variance level that is $\sim$40\% lower than the max-$\mathcal{C}$ metric, and comparable with the mini-$\mathcal{V}$ metric while not dropping the contrast. This indicates that even data with fewer measurements can be used to train the model and result in a considerable reduction in readout error, while never reducing the contrast. 
                     
Since there are only two examples in the training sets above, a large number of measurements, like $1\times 10^9$, for a training set alone may be specially needed to reduce the noise of the training set itself and make it more reliable. In fact, in NV-center experiments, the Rabi oscillations are always measured together with  optically detected magnetic resonance (ODMR) signal in priority, to extract feature information in preparation for the following experiments. The traditional method sums the photons at each microwave-pulse duration and discards the time-arrival information, but that information could actually be recorded. Therefore, we can employ the whole Rabi data as a training set to increase training examples, while the measurement repetitions can be fewer.  
 
\begin{figure}[htb]
\centering
\includegraphics[width=0.48\textwidth]{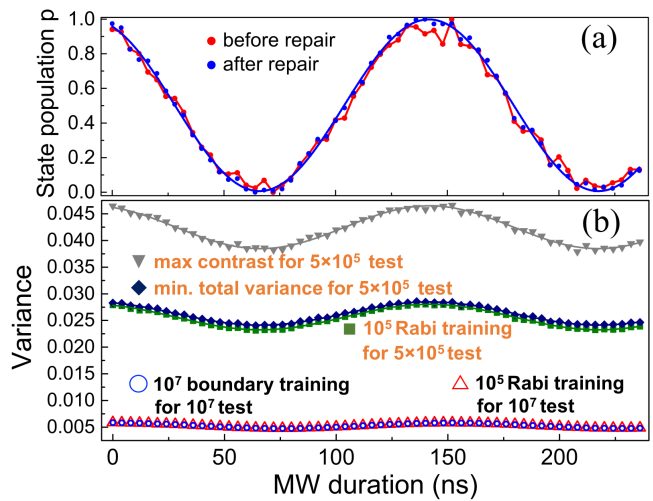}
\caption{ (a) Normalized original and repaired Rabi oscillations of  $5\times10^5$ measurements with a Rabi training set of $10^5$ measurements. (b) Results of processing Rabi oscillations with various methods. For the test set of $5\times10^5$ measurements, ML method training with $10^5$ Rabi data leads to 39.6\% lower in average variance than that of max-$\mathcal{C}$ metric, and 2.6\% lower than mini-$\mathcal{V}$ metric. For the $10^7$ test set, $10^5$  Rabi training is equivalent with $10^7$ boundary training. The Rabi training method is more efficient than the boundary training.}
\label{fig:rabi}
\end{figure} 
       
In Fig.\ref{fig:rabi}, we use a set of Rabi data measured with  $10^5$ repetitions  to train the model. For the peak point and the trough point, we assign with target values $q=1$ and $q=0$, respectively. For the rest of the examples, each target value is confirmed between 0 and 1 through  a sinusoidal fitting of the Rabi oscillations. After applying the model to another test set of Rabi data with $5\times10^5$ measurements, the optimized Rabi points are better aligned on a sinusoidal curve than the original data and show a slightly improved contrast, as shown in Fig.\ref{fig:rabi}(a). This indicates that the trained model repaired the result from imperfect test data and recovered more information damaged by noise. We believe that the repair effect will be better if noisier test data of fewer measurements, like $10^3$, are used.  We plot the variances of this repaired test set in Fig.\ref{fig:rabi}(b), together with the variances in the two traditional metrics. Here the Rabi training set of only $10^5$ measurements can result in 39.6\% lower in average variance than that of max-$\mathcal{C}$ metric, and 2.6\% lower than mini-$\mathcal{V}$ metric, while not reducing the contrast at all. On the other hand, we compared the processing effects on a test set of $10^7$ measurements, with a $10^5$ Rabi training set and a $10^7$ boundary training set, and find they are equivalent. Therefore, the Rabi training way is much more efficient than the boundary training way because it consumes no additional experimental time, while the calculation time with computer is not a concern. 
       
In summary, we demonstrate the application of ML to the spin readout of room-temperature NV center in diamond. We use linear regression technique to learn from fluorescence time traces of different spin states and build up a regression coefficient model, which is then applied to the processing of Rabi test data. The results show a marked reduction in readout error of state population without reducing the contrast compared with the traditional method. Our ML method can repair the result from imperfect data  and maximize the readout of information from noise. It can be concluded that the ML method raises the basic level of spin readout performance with data processing alone. This improvement will raise the sensitivity of many NV-center applications such as magnetometry, thermometry and electric field sensor. The ML method is easily extendable to other systems where the discrimination of states is basically relying on the readout of time-resolvable fluorescence via single-photon detection schemes\cite{PhysRevLett.117.013602, PhysRevA.97.013806}, such as the spin qubit readout in single quantum dot\cite{delteil2014observation}, and silicon-vacancy center in diamond\cite{sukachev2017silicon}. Moreover, in the fields involving fluorescence spectroscopy in biomedical engineering, our ML learning method might be employed to help distinguishing different enzymatic reaction steps\cite{terentyeva2013time}, and malignant tumors from non-malignant tissues in cancer detection\cite{alfano2012advances}.
    
This work is supported by the National Natural Science Foundation of China (Grant Nos. 11904070, 11604069 and 61805064), the National Key R\&D Program of China (Grant Nos. 2020YFA0309400, 2018YFA0306600 and 2018YFF01012500) and the Fundamental Research Funds for the Central Universities (No. PA2019GDQT0023).

\textbf{Data Availability} The data that support the findings of this study are available from the corresponding author
upon reasonable request.

\bibliography{ML}

\end{document}